



\font\twelverm=cmr10 scaled 1200    \font\twelvei=cmmi10 scaled 1200
\font\twelvesy=cmsy10 scaled 1200   \font\twelveex=cmex10 scaled 1200
\font\twelvebf=cmbx10 scaled 1200   \font\twelvesl=cmsl10 scaled 1200
\font\twelvett=cmtt10 scaled 1200   \font\twelveit=cmti10 scaled 1200

\skewchar\twelvei='177   \skewchar\twelvesy='60


\def\twelvepoint{\normalbaselineskip=12.4pt
  \abovedisplayskip 12.4pt plus 3pt minus 9pt
  \belowdisplayskip 12.4pt plus 3pt minus 9pt
  \abovedisplayshortskip 0pt plus 3pt
  \belowdisplayshortskip 7.2pt plus 3pt minus 4pt
  \smallskipamount=3.6pt plus1.2pt minus1.2pt
  \medskipamount=7.2pt plus2.4pt minus2.4pt
  \bigskipamount=14.4pt plus4.8pt minus4.8pt
  \def\rm{\fam0\twelverm}          \def\it{\fam\itfam\twelveit}%
  \def\sl{\fam\slfam\twelvesl}     \def\bf{\fam\bffam\twelvebf}%
  \def\mit{\fam 1}                 \def\cal{\fam 2}%
  \def\tt{\twelvett}
  \textfont0=\twelverm   \scriptfont0=\tenrm   \scriptscriptfont0=\sevenrm
  \textfont1=\twelvei    \scriptfont1=\teni    \scriptscriptfont1=\seveni
  \textfont2=\twelvesy   \scriptfont2=\tensy   \scriptscriptfont2=\sevensy
  \textfont3=\twelveex   \scriptfont3=\twelveex
\scriptscriptfont3=\twelveex
  \textfont\itfam=\twelveit
  \textfont\slfam=\twelvesl
  \textfont\bffam=\twelvebf \scriptfont\bffam=\tenbf
  \scriptscriptfont\bffam=\sevenbf
  \normalbaselines\rm}



\def\beginlinemode{\endmode
  \begingroup\parskip=0pt
\obeylines\def\\{\par}\def\endmode{\par\endgroup}}
\def\beginparmode{\endmode
  \begingroup \def\endmode{\par\endgroup}}
\let\endmode=\par
{\obeylines\gdef\
{}}
\def\singlespace{\baselineskip=\normalbaselineskip}

\def\oneandahalfspace{\baselineskip=\normalbaselineskip
  \multiply\baselineskip by 3 \divide\baselineskip by 2}
\def\doublespace{\baselineskip=\normalbaselineskip \multiply\baselineskip
by 2}

\newcount\firstpageno
\firstpageno=2
\footline={\ifnum\pageno<\firstpageno{\hfil}%
\else{\hfil\twelverm\folio\hfil}\fi}
\let\rawfootnote=\footnote              
\def\footnote#1#2{{\rm\singlespace\parindent=0pt\rawfootnote{#1}{#2}}}
\def\raggedcenter{\leftskip=4em plus 12em \rightskip=\leftskip
  \parindent=0pt \parfillskip=0pt \spaceskip=.3333em \xspaceskip=.5em
  \pretolerance=9999 \tolerance=9999
  \hyphenpenalty=9999 \exhyphenpenalty=9999 }
\def\dateline{\rightline{\ifcase\month\or
  January\or February\or March\or April\or May\or June\or
  July\or August\or September\or October\or November\or December\fi
  \space\number\year}}
\def\received{\vskip 3pt plus 0.2fill
 \centerline{\sl (Received\space\ifcase\month\or
  January\or February\or March\or April\or May\or June\or
  July\or August\or September\or October\or November\or December\fi
  \qquad, \number\year)}}


\hsize=6.5truein
\vsize=8.9truein
\parskip=\medskipamount
\twelvepoint            
\doublespace            
\overfullrule=0pt       


\def\preprintno#1{
 \rightline{\rm #1}}    

\def\title                      
  {\null\vskip 3pt plus 0.2fill
   \beginlinemode \doublespace \raggedcenter \bf}

\def\author                     
  {\vskip 3pt plus 0.2fill \beginlinemode
   \singlespace \raggedcenter}

\def\affil                      
  {\vskip 3pt plus 0.1fill \beginlinemode
   \oneandahalfspace \raggedcenter \sl}

\def\abstract                   
  {\vskip 3pt plus 0.3fill \beginparmode
   \doublespace \narrower ABSTRACT: }

\def\endtitlepage               
  {\endpage                     
   \body}

\def\body                       
  {\beginparmode}               

\def\subhead#1{                 
  \vskip 0.25truein             
  {\raggedcenter #1 \par}
   \nobreak\vskip 0.25truein\nobreak}

\def\refto#1{[{#1}]}           

\def\references                 
  {\subhead{References}   
   \beginparmode
   \frenchspacing \parindent=0pt \leftskip=1truecm
   \parskip=8pt plus 3pt \everypar{\hangindent=\parindent}}

\gdef\refis#1{\indent\hbox to 0pt{\hss#1.~}}    

\gdef\journal#1, #2, #3, 1#4#5#6{               
    {\sl #1~}{\bf #2}, #3, (1#4#5#6)}        

\def\refstylenp{                
  \gdef\refto##1{ [##1]}                                
  \gdef\refis##1{\indent\hbox to 0pt{\hss##1)~}}        
  \gdef\journal##1, ##2, ##3, ##4 {                     
     {\sl ##1~}{\bf ##2~}(##3) ##4 }}

\def\refstyleprnp{              
  \gdef\refto##1{ [##1]}                                
  \gdef\refis##1{\indent\hbox to 0pt{\hss##1)~}}        
  \gdef\journal##1, ##2, ##3, 1##4##5##6{               
    {\sl ##1~}{\bf ##2~}(1##4##5##6) ##3}}

\def\pr{\journal Phys. Rev., }

\def\prd{\journal Phys. Rev. D, }

\def\prl{\journal Phys. Rev. Lett., }

\def\jmp{\journal J. Math. Phys., }

\def\rmp{\journal Rev. Mod. Phys., }

\def\cqg{\journal Class. Quantum Grav., }

\def\cmp{\journal Comm. Math. Phys., }

\def\np{\journal Nucl. Phys., }

\def\endreferences{\body}

\def\figurecaptions             
  { \beginparmode
   \subhead{Figure Captions}
}

\def\endpage                    
  {\vfill\eject}

\def\endpaper                   
  {\endmode\vfill\supereject}

\def\endit
  {\endpaper\end}


\def\ref#1{Ref. #1}                     
\def\Ref#1{Ref. #1}                     

\def\frac#1#2{{\textstyle{#1 \over #2}}}

\def\ie{{\it i.e.,\ }}

\def\sla{\raise.15ex\hbox{$/$}\kern-.57em}
\def\leaderfill{\leaders\hbox to 1em{\hss.\hss}\hfill}
\def\twiddle{\lower.9ex\rlap{$\kern-.1em\scriptstyle\sim$}}
\def\bigtwiddle{\lower1.ex\rlap{$\sim$}}
\def\gtwid{\mathrel{\raise.3ex\hbox{$>$\kern-
.75em\lower1ex\hbox{$\sim$}}}}
\def\ltwid{\mathrel{\raise.3ex\hbox{$<$\kern-
.75em\lower1ex\hbox{$\sim$}}}}
\def\square{\kern1pt\vbox{\hrule height 1.2pt\hbox{\vrule width
1.2pt\hskip 3pt
   \vbox{\vskip 6pt}\hskip 3pt\vrule width 0.6pt}\hrule height
0.6pt}\kern1pt}

\def\m@th{\mathsurround=0pt }
\def\leftrightarrowfill{$\m@th \mathord\leftarrow \mkern-6mu
 \cleaders\hbox{$\mkern-2mu \mathord- \mkern-2mu$}\hfill
 \mkern-6mu \mathord\rightarrow$}
\def\overleftrightarrow#1{\vbox{\ialign{##\crcr
     \leftrightarrowfill\crcr\noalign{\kern-1pt\nointerlineskip}
     $\hfil\displaystyle{#1}\hfil$\crcr}}}


\font\titlefont=cmr10 scaled\magstep3

\def\martinstyletitle                      
  {\null\vskip 3pt plus 0.2fill
   \beginlinemode \doublespace \raggedcenter \titlefont}

\font\twelvesc=cmcsc10 scaled 1200

\def\author                     
  {\vskip 3pt plus 0.2fill \beginlinemode
   \singlespace \raggedcenter\twelvesc}


\def\heading                            
  {\vskip 0.5truein plus 0.1truein      
\endheading
   \beginparmode \def\\{\par} \parskip=0pt \singlespace \raggedcenter}

\def\endheading
  {\par\nobreak\vskip 0.25truein\nobreak\beginparmode}

\def\subheading                         
  {\vskip 0.25truein plus 0.1truein     
   \beginlinemode \singlespace \parskip=0pt \def\\{\par}\raggedcenter}

\def\tag#1$${\eqno(#1)$$}

\def\align#1$${\eqalign{#1}$$}

\def\aligntag#1$${\gdef\tag##1\\{&(##1)\cr}\eqalignno{#1\\}$$
  \gdef\tag##1$${\eqno(##1)$$}}

\def\endaligntag{}

\def\overset #1\to#2{{\mathop{#2}\limits^{#1}}}
\def\underset#1\to#2{{\let\next=#1\mathpalette\undersetpalette#2}}
\def\undersetpalette#1#2{\vtop{\baselineskip0pt
\ialign{$\mathsurround=0pt #1\hfil##\hfil$\crcr#2\crcr\next\crcr}}}


\def\ref#1{Ref.~#1}                     
\def\Ref#1{Ref.~#1}                     
\def\[#1]{[\cite{#1}]}
\def\cite#1{{#1}}
\def\(#1){(\call{#1})}
\def\call#1{{#1}}
\def\taghead#1{}
\def\frac#1#2{{#1 \over #2}}

\def\12{{1\over2}}

\def\ie{{\it i.e.,\ }}

\def\sla{\raise.15ex\hbox{$/$}\kern-.57em}
\def\leaderfill{\leaders\hbox to 1em{\hss.\hss}\hfill}
\def\twiddle{\lower.9ex\rlap{$\kern-.1em\scriptstyle\sim$}}
\def\bigtwiddle{\lower1.ex\rlap{$\sim$}}
\def\gtwid{\mathrel{\raise.3ex\hbox{$>$\kern-
.75em\lower1ex\hbox{$\sim$}}}}
\def\ltwid{\mathrel{\raise.3ex\hbox{$<$\kern-
.75em\lower1ex\hbox{$\sim$}}}}
\def\square{\kern1pt\vbox{\hrule height 1.2pt\hbox{\vrule width
1.2pt\hskip 3pt
   \vbox{\vskip 6pt}\hskip 3pt\vrule width 0.6pt}\hrule height
0.6pt}\kern1pt}
\def\tdot#1{\mathord{\mathop{#1}\limits^{\kern2pt\ldots}}}

\def\pmb#1{\setbox0=\hbox{#1}%
  \kern-.025em\copy0\kern-\wd0
  \kern  .05em\copy0\kern-\wd0
  \kern-.025em\raise.0433em\box0 }

\catcode`@=11
\newcount\tagnumber\tagnumber=0

\immediate\newwrite\eqnfile
\newif\if@qnfile\@qnfilefalse
\def\write@qn#1{}
\def\writenew@qn#1{}
\def\w@rnwrite#1{\write@qn{#1}\message{#1}}
\def\@rrwrite#1{\write@qn{#1}\errmessage{#1}}

\def\taghead#1{\gdef\t@ghead{#1}\global\tagnumber=0}
\def\t@ghead{}

\expandafter\def\csname @qnnum-3\endcsname
  {{\t@ghead\advance\tagnumber by -3\relax\number\tagnumber}}
\expandafter\def\csname @qnnum-2\endcsname
  {{\t@ghead\advance\tagnumber by -2\relax\number\tagnumber}}
\expandafter\def\csname @qnnum-1\endcsname
  {{\t@ghead\advance\tagnumber by -1\relax\number\tagnumber}}
\expandafter\def\csname @qnnum0\endcsname
  {\t@ghead\number\tagnumber}
\expandafter\def\csname @qnnum+1\endcsname
  {{\t@ghead\advance\tagnumber by 1\relax\number\tagnumber}}
\expandafter\def\csname @qnnum+2\endcsname
  {{\t@ghead\advance\tagnumber by 2\relax\number\tagnumber}}
\expandafter\def\csname @qnnum+3\endcsname
  {{\t@ghead\advance\tagnumber by 3\relax\number\tagnumber}}

\def\equationfile{%
  \@qnfiletrue\immediate\openout\eqnfile=\jobname.eqn%
  \def\write@qn##1{\if@qnfile\immediate\write\eqnfile{##1}\fi}
  \def\writenew@qn##1{\if@qnfile\immediate\write\eqnfile
    {\noexpand\tag{##1} = (\t@ghead\number\tagnumber)}\fi}
}

\def\callall#1{\xdef#1##1{#1{\noexpand\call{##1}}}}
\def\call#1{\each@rg\callr@nge{#1}}

\def\each@rg#1#2{{\let\thecsname=#1\expandafter\first@rg#2,\end,}}
\def\first@rg#1,{\thecsname{#1}\apply@rg}
\def\apply@rg#1,{\ifx\end#1\let\next=\relax%
\else,\thecsname{#1}\let\next=\apply@rg\fi\next}

\def\callr@nge#1{\calldor@nge#1-\end-}
\def\callr@ngeat#1\end-{#1}
\def\calldor@nge#1-#2-{\ifx\end#2\@qneatspace#1 %
  \else\calll@@p{#1}{#2}\callr@ngeat\fi}
\def\calll@@p#1#2{\ifnum#1>#2{\@rrwrite{Equation range #1-#2\space
is bad.}
\errhelp{If you call a series of equations by the notation M-N, then M and
N must be integers, and N must be greater than or equal to M.}}\else %
{\count0=#1\count1=#2\advance\count1
by1\relax\expandafter\@qncall\the\count0,%
  \loop\advance\count0 by1\relax%
    \ifnum\count0<\count1,\expandafter\@qncall\the\count0,%
  \repeat}\fi}

\def\@qneatspace#1#2 {\@qncall#1#2,}
\def\@qncall#1,{\ifunc@lled{#1}{\def\next{#1}\ifx\next\empty\else
  \w@rnwrite{Equation number \noexpand\(>>#1<<) has not been defined
yet.}
  >>#1<<\fi}\else\csname @qnnum#1\endcsname\fi}

\let\eqnono=\eqno
\def\eqno(#1){\tag#1}
\def\tag#1$${\eqnono(\displayt@g#1 )$$}

\def\aligntag#1\endaligntag
  $${\gdef\tag##1\\{&(##1 )\cr}\eqalignno{#1\\}$$
  \gdef\tag##1$${\eqnono(\displayt@g##1 )$$}}

\def\eqalignno#1{\displ@y \tabskip\centering
  \halign to\displaywidth{\hfil$\displaystyle{##}$\tabskip\z@skip
    &$\displaystyle{{}##}$\hfil\tabskip\centering
    &\llap{$\displayt@gpar##$}\tabskip\z@skip\crcr
    #1\crcr}}

\def\displayt@gpar(#1){(\displayt@g#1 )}

\def\displayt@g#1 {\rm\ifunc@lled{#1}\global\advance\tagnumber by1
        {\def\next{#1}\ifx\next\empty\else\expandafter
        \xdef\csname
@qnnum#1\endcsname{\t@ghead\number\tagnumber}\fi}%
  \writenew@qn{#1}\t@ghead\number\tagnumber\else
        {\edef\next{\t@ghead\number\tagnumber}%
        \expandafter\ifx\csname @qnnum#1\endcsname\next\else
        \w@rnwrite{Equation \noexpand\tag{#1} is a duplicate
number.}\fi}%
  \csname @qnnum#1\endcsname\fi}

\def\ifunc@lled#1{\expandafter\ifx\csname @qnnum#1\endcsname\relax}

\let\@qnend=\end\gdef\end{\if@qnfile
\immediate\write16{Equation numbers written on
[]\jobname.EQN.}\fi\@qnend}

\catcode`@=12

\catcode`@=11
\newcount\r@fcount \r@fcount=0
\newcount\r@fcurr
\immediate\newwrite\reffile
\newif\ifr@ffile\r@ffilefalse
\def\w@rnwrite#1{\ifr@ffile\immediate\write\reffile{#1}\fi\message{#1}}

\def\writer@f#1>>{}
\def\referencefile{
  \r@ffiletrue\immediate\openout\reffile=\jobname.ref%
  \def\writer@f##1>>{\ifr@ffile\immediate\write\reffile%
    {\noexpand\refis{##1} = \csname r@fnum##1\endcsname = %
     \expandafter\expandafter\expandafter\strip@t\expandafter%
     \meaning\csname r@ftext\csname
r@fnum##1\endcsname\endcsname}\fi}%
  \def\strip@t##1>>{}}

\def\citeall#1{\xdef#1##1{#1{\noexpand\cite{##1}}}}
\def\cite#1{\each@rg\citer@nge{#1}}	

\def\each@rg#1#2{{\let\thecsname=#1\expandafter\first@rg#2,\end,}}
\def\first@rg#1,{\thecsname{#1}\apply@rg}	
\def\apply@rg#1,{\ifx\end#1\let\next=\relax
\else,\thecsname{#1}\let\next=\apply@rg\fi\next}

\def\citer@nge#1{\citedor@nge#1-\end-}	
\def\citer@ngeat#1\end-{#1}
\def\citedor@nge#1-#2-{\ifx\end#2\r@featspace#1 
  \else\citel@@p{#1}{#2}\citer@ngeat\fi}	
\def\citel@@p#1#2{\ifnum#1>#2{\errmessage{Reference range #1-
#2\space is bad.}%
    \errhelp{If you cite a series of references by the notation M-N, then M
and
    N must be integers, and N must be greater than or equal to M.}}\else%
 {\count0=#1\count1=#2\advance\count1
by1\relax\expandafter\r@fcite\the\count0,
  \loop\advance\count0 by1\relax
    \ifnum\count0<\count1,\expandafter\r@fcite\the\count0,%
  \repeat}\fi}

\def\r@featspace#1#2 {\r@fcite#1#2,}	
\def\r@fcite#1,{\ifuncit@d{#1}
    \newr@f{#1}%
    \expandafter\gdef\csname r@ftext\number\r@fcount\endcsname%
                     {\message{Reference #1 to be supplied.}%
                      \writer@f#1>>#1 to be supplied.\par}%
 \fi%
 \csname r@fnum#1\endcsname}
\def\ifuncit@d#1{\expandafter\ifx\csname r@fnum#1\endcsname\relax}%
\def\newr@f#1{\global\advance\r@fcount by1%
    \expandafter\xdef\csname r@fnum#1\endcsname{\number\r@fcount}}

\let\r@fis=\refis			
\def\refis#1#2#3\par{\ifuncit@d{#1}
blank
   \newr@f{#1}%
   \w@rnwrite{Reference #1=\number\r@fcount\space is not cited up to
now.}\fi%
  \expandafter\gdef\csname r@ftext\csname
r@fnum#1\endcsname\endcsname%
  {\writer@f#1>>#2#3\par}}

\def\ignoreuncited{
   \def\refis##1##2##3\par{\ifuncit@d{##1}%
    \else\expandafter\gdef\csname r@ftext\csname
r@fnum##1\endcsname\endcsname%
     {\writer@f##1>>##2##3\par}\fi}}

\def\r@ferr{\endreferences\errmessage{I was expecting to see
\noexpand\endreferences before now;  I have inserted it here.}}
\let\r@ferences=\references
\def\references{\r@ferences\def\endmode{\r@ferr\par\endgroup}}

\let\endr@ferences=\endreferences
\def\endreferences{\r@fcurr=0
  {\loop\ifnum\r@fcurr<\r@fcount
    \advance\r@fcurr by
1\relax\expandafter\r@fis\expandafter{\number\r@fcurr}%
    \csname r@ftext\number\r@fcurr\endcsname%
  \repeat}\gdef\r@ferr{}\endr@ferences}


\let\r@fend=\endpaper\gdef\endpaper{\ifr@ffile
\immediate\write16{Cross References written on
[]\jobname.REF.}\fi\r@fend}

\catcode`@=12

\citeall\refto		
\citeall\ref		%
\citeall\Ref		%
\ignoreuncited
\preprintno{FTG-116-USU, 6/93}
\doublespace
\title Gravitational Observables and Local Symmetries
\author C. G. Torre
\affil Department of Physics
Utah State University
Logan, UT  84322-4415
USA
\abstract
Using a recent classification of local symmetries of the vacuum Einstein
equations, it is shown that there can be no observables for the vacuum
gravitational field (in a closed universe) built as spatial integrals of local
functions of Cauchy data
and their derivatives.
\endtitlepage

A long-standing open problem in Einstein's general theory of relativity is
to give an invariant characterization of the state of the (vacuum)
gravitational field in terms of quantities measurable at a single instant of
time.  Finding such a characterization constitutes the well-known ``problem
of observables'' in Hamiltonian relativity \refto{Bergmann1961}.  A
precise formulation of
this problem is as follows.  Let $\Gamma$ denote the phase space for
general relativity.  To fix ideas, let us choose $\Gamma$ to be the
cotangent bundle over the space of Riemannian metrics on a compact
three-dimensional manifold $\Sigma$.  A point in phase space can be fixed
by specifying a pair $(q_{ab},p^{ab})$, where $q_{ab}$ is a metric on
$\Sigma$ and $p^{ab}$ is a symmetric tensor density on $\Sigma$.  A
point $x\in\Gamma$ defines a
state of
the gravitational field if and only if it lies in the subspace
$\overline\Gamma\subset\Gamma$ defined as the locus of points satisfying
the Hamiltonian and momentum constraints \refto{York1980,Fischer1979}
$$
{\cal H}=0={\cal H}_a.\tag1a
$$
$\cal H$ and ${\cal H}_a$ are often called the ``super-Hamiltonian'' and
``super-momentum''.
Viewing the constraints as
vanishing of functions on $\Gamma$ we can express them as
$$
H(N)=0=H({\bf N}) \hskip0.5truein\forall N, {\bf N}.\tag1
$$
Here $H(N)$ is the super-Hamiltonian
smeared with a ``lapse function'', which is any function on $\Sigma$;
$H({\bf N})$ is the super-momentum
smeared with a vector field on $\Sigma$, often called the ``shift vector'':
$$
\eqalign{H(N)&=\int_\Sigma N {\cal H}\cr
H({\bf N})&=\int_\Sigma N^a {\cal H}_a.}
$$

While each point of $\overline\Gamma$ defines a gravitational field, the
description is rather redundant:  infinitely many points in
$\overline\Gamma$ define the same gravitational field
\refto{Bergmann1961}.  As is well
known, for each point $x\in\overline\Gamma$ and for each choice of lapse
and shift there is a 1-parameter family of
points on $\overline\Gamma$ that are physically equivalent to $x$.  This
curve of redundancy is the flow through $x$
 of the Hamiltonian vector field
defined by the constraint function $H(N)+H({\bf N})$.
For each $N$ and $\bf N$, $H(N)+H({\bf N})$ represents a
Hamiltonian for the Einstein equations, so the flow connecting
physically equivalent canonical data represents time evolution.
Infinitesimally, $H(N)$ generates the canonical transformation
of the phase space data induced by a normal
deformation of $\Sigma$ (now thought of as embedded in the Einstein
space) specified at each
point by $N$.  Similarly, $H({\bf N})$ provides the infinitesimal canonical
transformation of the data induced by a tangential
deformation of $\Sigma$ specified by $\bf N$.  Normal and
tangential deformations of the hypersurface can be viewed as
the action on the hypersurface of infinitesimal diffeomorphisms of the
spacetime manifold $\cal M$. The corresponding
canonical transformations represent the change in the canonical data as they
are carried by the (infinitesimal) diffeomorphism from point to point in
the Einstein space
for which they are the Cauchy data.

{}From the above discussion it is clear that a non-redundant characterization
of the state of the gravitational field involves finding functions on
$\overline\Gamma$ invariant under the flow generated by $H(N)+H({\bf
N})$.  Such ``observables'' are functions of Cauchy data that are invariant
under infinitesimal spacetime diffeomorphisms modulo the Einstein
equations.  More succinctly, the observables are constants of motion for the
Einstein equations.  A mathematical characterization is easily found:  the
observables are equivalence classes of functions $F:\Gamma\to R$ that have
weakly vanishing
Poisson brackets with the constraint functions $H(N)$, $H({\bf N})$ for all
$N$ and $\bf N$:
$$
[F, H(N)]\bigg|_{\overline\Gamma}=0=[F, H({\bf
N})]\bigg|_{\overline\Gamma}.\tag2
$$
Two functions $F_1$ and $F_2$ are equivalent if their difference vanishes
on $\overline\Gamma$:
$$
F_1\sim F_2 \Longleftrightarrow (F_1-F_2)\bigg|_{\overline\Gamma}=0.
$$

If $\Sigma$ is open, and asymptotically flat boundary conditions are
included in the definition of $\Gamma$, then the ADM energy,
momentum,
and angular momentum provide examples of observables.  Clearly this
handful of constants of motion is inadequate to characterize completely
the state of the gravitational field.  If $\Sigma$ is compact without
boundary there are {\it no known observables}.  In the classical
theory the scarcity of known observables is perhaps only a technical
annoyance.   This annoyance
becomes a stumbling block when the rules of Dirac constraint quantization
are applied to construct a quantum theory of gravity
\refto{Ashtekar1991a}.  Here
observables play a key role, and their scarcity hampers progress in
quantum gravity.  Here we will show that the complexity of the
Einstein equations prohibits the simplest class of putative observables from
existing.  Henceforth, unless otherwise stated, we will assume the universe
is closed, \ie $\Sigma$ is compact without boundary.

If one could integrate the Einstein equations and find an internal time, then
in principle a complete set of observables could be found
\refto{CGT1991c}.
Unfortunately, it is unlikely that the general solution of the Einstein
equations will be available any time soon, and it is quite problematic to
isolate internal spacetime variables from $\Gamma$ \refto{CGT1992a}.
A direct
systematic search for observables would seem to be intractable if only
because of the bewildering array of ways to attempt their construction.
Nevertheless, let us begin such a search.
The simplest class of functions on $\Gamma$ that one can consider are the
{\it local functionals}, built as integrals over $\Sigma$ of local functions of
the canonical variables $(q_{ab},p^{ab})$ and their derivatives.  By ``local
functions'' is meant
that at a given point $x\in\Sigma$ the function being integrated depends on
the canonical variables and their derivatives up to some finite order {\it at
$x$}.  For example, the constraint functions $H(N)$ and $H({\bf N})$ are
local functionals; they are observables too, but they are equivalent to zero.
In the asymptotically flat context the energy, momentum and
angular momentum observables can be
viewed as local functionals.  So we would like to answer the question:  Are
there any (non-trivial) observables for closed universes built as local
functionals of the
canonical data?  As we shall see, the answer is no.  The key to showing this
is to use the fact that if a local functional is an
observable, then there must be a corresponding local ``hidden symmetry''
for the Einstein equations.

Let $F:\Gamma\to
R$ be such an observable.  Because it is a local functional, we can
rigorously assert that \(2) is equivalent to the following Poisson bracket
relations
$$
[F,{\cal H}_\alpha(x)] = \int_\Sigma dy\, \Lambda_\alpha^\beta(x,y) {\cal
H}_\beta(y),
$$
where $\Lambda_\alpha^\beta(x,y)$ is built from local functions of the
canonical variables, delta functions and derivatives of delta functions to
some finite order; we have defined ${\cal H}_\alpha=({\cal H},{\cal
H}_a)$.
Corresponding to $F$ is the Hamiltonian vector field $V_F$  defined by
$$
V_F = \int_\Sigma\left( {\delta F\over\delta p^{ab}} {\delta \over\delta
q_{ab}}-{\delta F\over\delta q_{ab}} {\delta \over\delta p^{ab}}\right).
$$
$V_F$ is the infinitesimal generator of a one parameter family of
canonical transformations mapping
admissible Cauchy data to other admissible data, \ie mapping
solutions at any given time to other solutions at that time.  Infinitesimally,
the canonical transformation is given by
$$
\eqalign{&\delta q_{ab} = V_F(q_{ab})={\delta F\over\delta p^{ab}}\cr
&\delta p^{ab} = V_F(p^{ab})=-{\delta F\over\delta q_{ab}}.}\tag5
$$
Because $F$ is a
local functional, the components of $V_F$ associated with the chart $(
q_{ab},p^{ab})$, given by ${\delta F\over\delta p^{ab}}$ and $-{\delta
F\over\delta q_{ab}}$, are local functions of the canonical variables.

Now, let $( q_{ab}(t), p^{ab}(t))$ denote a solution to the Hamilton
equations for a given choice of lapse and shift $N^\alpha=(N(t),{\bf
N}(t))$.   This means that, at each $t$, $( q_{ab}(t), p^{ab}(t))$ satisfy the
constraints \(1a) and the evolution equations defined by the Hamiltonian
$H(N)+H({\bf N})$.  Because of the requirement \(2), the infinitesimal
transformation $(\delta q_{ab}, \delta p^{ab}, \delta N^\alpha)$, given by
\(5) and
$$
\delta N^\alpha(y)=\int_\Sigma dx
N^\beta(x)\Lambda_\beta^\alpha(x,y),\tag6
$$
satisfies the Hamilton equations linearized about the solution $( q_{ab}(t),
p^{ab}(t),N^\alpha(t))$.

The spacetime metric $g_{ab}$ which solves the Einstein equations is
constructed algebraically from $ q_{ab}(t)$ and $N^\alpha(t)$.
Conversely, given a spacetime Einstein metric, one can reconstruct the one
parameter family $( q_{ab}(t), p^{ab}(t), N^\alpha(t))$ algebraically (and
hence locally) from the spacetime metric and its first derivatives
\refto{reconstruct}.  Note in particular that, in a solution to the Hamilton
equations, the canonical momentum $p^{ab}(t)$ is constructed
algebraically from the 3-metric, the lapse and shift, and their first
derivatives.
Therefore, the infinitesimal transformation generated by $F$
will correspond to a change $\delta g_{ab}$ in the spacetime metric that is
a local
function of $g_{ab}$ and a finite number of
its derivatives at a point. It is straightforward to see that $\delta g_{ab}$
satisfies the spacetime form of the Einstein equations linearized about
$g_{ab}$. In this fashion the observable generates an infinitesimal map of
solutions to solutions.  Local transformations of this type mapping
solutions to solutions are called ``generalized symmetries'' by
mathematicians.

Recently all generalized symmetries of the vacuum
Einstein equations have been classified \refto{Symmetries}.  They consist
of a trivial scaling
symmetry and the familiar diffeomorphism symmetry.  The former cannot
be implemented as a symplectic map of $\Gamma$, while the latter is
generated by the constraint functions themselves.  Because there are no
other symmetries, there can be no observables (save the trivial constraints)
built as local functionals of the canonical variables.

A more explicit proof of this relies on the connection between symmetries
and conservation laws.  An observable $F$ that is built as a local functional
corresponds to a local differential conservation law, \ie a spacetime 3-form
$\sigma$ that is closed by virtue of the Einstein equations.   To see this we
first note that $F$ is, by definition, an integral over $\Sigma$ of a spatial
3-form $\hat\sigma$ built locally from $x\in\Sigma$, the canonical
variables $( q_{ab},p^{ab})$ and their derivatives:
$$
F[q,p] = \int_\Sigma\hat\sigma(x,q,p,\partial_x q,\partial_x p,\dots),\tag3a
$$
Because of \(2), if we evaluate $\hat\sigma$ on any solution $(
q_{ab}(t),p^{ab}(t))$ then $F$ is independent of $t$.  This will be true for
solutions constructed using any lapse and shift.  As before, we can translate
this result into spacetime form in terms of the Einstein metric $g_{ab}$
defined by $( q_{ab}(t),N(t), {\bf N}(t))$ .  From this point of view we
obtain from $\hat\sigma$ a spacetime 3-form $\sigma(x,g,\partial_x
g,\dots)$ built locally from $x\in {\cal M}$,  the spacetime metric and its
derivatives.  We thus obtain a functional of $g_{ab}$ via
$$
F[g] = \int_\Sigma\sigma(x,g,\partial_x g,\dots),\tag3
$$
where now $\Sigma$ is viewed as a spacelike hypersurface rather than an
abstract 3-manifold, and we have for simplicity used the same symbol $F$
to denote the resulting functional of the spacetime metric.  Eq. \(2) implies
that the value of $F[g]$ is independent of the choice of $\Sigma$ when
$g_{ab}$ satisfies the vacuum Einstein equations.  Therefore the exterior
derivative of $\sigma$ vanishes when the Einstein equations are satisfied.

As a byproduct of the symmetry classification of \refto{Symmetries} it was
shown that all
weakly closed 3-forms are weakly equivalent to identically (\ie strongly)
closed 3-forms \refto{weakly}.   Thus, because of the trivial nature of the
symmetries of the vacuum Einstein equations,  local conservation laws are
essentially topological in nature.  The proper setting for understanding this
is the variational bicomplex \refto{Anderson1992} associated with the jet
bundle of metrics over spacetime.  In that context it can be shown that an
identically closed 3-form $\sigma$ built locally from the spacetime metric
and its derivatives (as well as the spacetime position)  can be written as the
sum of an exact
form and a representative $\sigma_{\scriptscriptstyle 0}$ of the
cohomology class of $\sigma$:
$$
\sigma = d\alpha + \sigma_{\scriptscriptstyle 0},\tag4
$$
where $\alpha$ and $\sigma_{\scriptscriptstyle 0}$ are also local functions
of the metric and its derivatives.
The relevant cohomology is the De Rham cohomology of the bundle of
metrics over spacetime.  We need not explore this cohomology here;
although the integral of $\sigma_{\scriptscriptstyle 0}$ over a
hypersurface {\it is} a constant of motion, it is a trivial one because this
functional of the metric is conserved irrespective of whether or not the
Einstein equations are satisfied.  Therefore only the 2-form $\alpha$ can
lead to non-trivial local observables.

In the asymptotically flat context, the structure of spatial infinity allows
non-trivial conservation laws, namely, that of energy, momentum and
angular momentum to be encoded in $\alpha$.  In detail, the integral of
$\sigma$ over $\Sigma$ involves an integral of $\alpha$
over the ``sphere at infinity'', and this leads to the ADM observables (for
appropriate choices of $\alpha$) \refto{Goldberg1980}.
If spacetime is diffeomorphic to $R\times\Sigma$ with $\partial\Sigma=0$
then no
asymptotic region can be used to construct non-trivial constants of motion
(built as local functionals) because now the integral over $d\alpha$ vanishes
identically.  In other words, for closed universes ``on
shell'',  the only possible
conservation laws derive from the topology of the bundle of metrics over
spacetime---this is the information contained in $\sigma_{\scriptscriptstyle
0}$---and
have nothing to do with the Einstein equations {\it per se\/}.  Thus there
can be no non-trivial observables for closed universes constructed as local
functionals.

It would seem then that observables must be constructed in a more
complicated fashion than a local functional.  Unfortunately, there does not
appear
to be any way of systematically identifying ``non-local conservation laws''
for the
Einstein equations.    In many examples non-local conservation laws for
partial differential
equations are closely tied to the integrability of those equations.  A
well-known attribute of an integrable system of partial differential
equations is
the existence of infinitely many generalized symmetries.  Modulo the
diffeomorphism symmetry, which is physically trivial, the Einstein
equations fail to pass this test and so one can expect little luck in finding
such non-local conservation laws based on some sort of integrability.
Indeed, there is a result of Kucha\v r
that rules out any observables built as linear functionals of the ADM
momenta \refto{Kuchar1981}.  One encouraging recent result
\refto{Lewandowski1992} shows that the
holonomy group of the Ashtekar connection on a given hypersurface is
almost a constant of motion.  For the meaning of ``almost'' see
\refto{Jacobson1992}.  Clearly
this type of observable is quite
non-local.  One can hope (but it is only a hope) that the results of
\refto{Lewandowski1992}
in the context of the Ashtekar canonical formalism are the hint of some
structure that can be used to find non-local conservation laws, at least in
principle.  In practice, it is possible that perturbative methods for defining
observables can be devised.  This is really an important possibility.  Given
the scarcity of exactly soluble quantum field theories, it is to be expected
that a quantum theory of gravity would need a perturbative definition
at some point.
So, while it seems possible to find the exact quantum states
\refto{Ashtekar1991a}, it may be
necessary to approximate the dynamical information contained in
the observables.  Hopefully, such a perturbation theory will be better
behaved than its weak-field counterpart.  Failing this, it appears that the
standard rules
for canonical quantization of constrained systems, in which the observables
play a central role, will have to be improved or modified to avoid the
problem of observables.

\noindent{\bf Acknowledgments}

It is a pleasure to thank Prof. Ian Anderson for helpful discussions.  This
work was supported in part by a Faculty Research Grant from Utah State
University.
\references\doublespace

\refis{reconstruct}{In order to construct an Einstein metric using the 1-
parameter family of canonical data, or to reconstruct the data from an
Einstein metric, one also needs to introduce a foliation of the spacetime
manifold $\cal M$.}

\refis{weakly}{By ``weakly'' we mean the relations hold modulo
the field equations.}

\refis{Symmetries}{C. G. Torre  and I. M. Anderson, \prl 70, 3525,
1993.}

\refis{Jacobson1992}{T. Jacobson and J. Romano, University of Maryland
preprint UMDGR-92-208, 1992.}

\refis{ADM1959}{R. Arnowitt, S. Deser, and C. Misner, \pr 116, 1322,
1959.}

\refis{ADM1962}{R. Arnowitt, S. Deser, and C. Misner in {\it
Gravitation: An
Introduction to Current Research}, edited by L. Witten (Wiley, New York
1962).}

\refis{Anderson1984}{I. M. Anderson, \journal Ann. Math., 120, 329,
1984.}

\refis{Anderson1992}{I. M. Anderson, ``Introduction to the Variational
Bicomplex'', in {\it Mathematical Aspects of Classical Field Theory} (Eds.
M. Gotay, J. Marsden, V. Moncrief),\journal Cont. Math., 32, 51, 1992.}

\refis{Arms1979}{J. Arms, \jmp 20, 443, 1979.}

\refis{Arms1980}{J. Arms, \jmp 21, 15, 1980.}

\refis{Ashtekar1991a}{A. Ashtekar, {\it Lectures on Non-Perturbative
Canonical
Gravity}, (World Scientific, Singapore 1991) and references therein.}

\refis{Ashtekar1991b}{ A. Ashtekar, L. Bombelli, and O.
Reula, in {\it Mechanics,
Analysis and Geometry : 200 Years After Lagrange}, edited by M.
Francaviglia ( North-Holland, New York 1991). }

\refis{Barrow1982}{J. Barrow, \prpts 85, 1, 1982.}

\refis{Belinsky1979}{V. Belinsky and V. Zakharov, \journal Sov. Phys.
JETP, 50, 1, 1979.}

\refis{Bergmann1961}{P. Bergmann, \rmp 33, 510, 1961.}

\refis{Bluman1989}{G. Bluman and S. Kumei, {\it Symmetries of
Differential Equations}, (Springer-Verlag, New York 1989).}

\refis{Choquet1982}{Y. Choquet-Bruhat, C. DeWitt-Morette, and M.
Dillard-Bleick, {\it Analysis, Manifolds and Physics}, (North-Holland,
New York 1982). }

\refis{Dirac1964}{P.A.M. Dirac, {\it Lectures on Quantum Mechanics},
(Yeshiva
University, New York, 1964).}

\refis{Epstein1975}{D. Epstein, \journal J. Diff. Geom., 10, 631, 1975.}

\refis{Fischer1979}{See A. Fischer and J. Marsden in {\it General
Relativity: An
Einstein Centenary Survey}, edited by S. Hawking and W. Israel
(Cambridge University Press, Cambridge 1979).}

\refis{Fokas1987}{A. Fokas, \journal Stud. Appl. Math., 77, 253, 1987.}

\refis{Goldberg1980}{J. Goldberg in {\it General
Relativity
and Gravitation: 100 Years After the Birth of Albert Einstein, Vol. 1},
edited by A. Held (Plenum, New York 1980).}

\refis{Halliwell1991}{J. Halliwell, \prd 43, 2590, 1991.}

\refis{Hauser1981}{I. Hauser and F. Ernst, \jmp 22, 1051, 1981.}

\refis{Henneaux1991}{G. Barnich, M. Henneaux, C. Schomblond, \prd 44,
R939, 1991.}

\refis{Hormander1966}{L. H\" ormander, \journal Ann. Math., 83, 129,
1966.}

\refis{Ibragimov1985}{N. Ibragimov, {\it Transformation Groups
Applied to Mathematical Physics}, (D. Reidel, Boston 1985).}

\refis{Isenberg1982}{J. Isenberg and J.
Marsden, \prpts 89, 181, 1982, and references therein.}

\refis{Isham1985}{C. J. Isham and K. V. Kucha\v r, \ann 164, 288, 1985;
\ann 164, 316, 1985.}

\refis{Kuchar1971}{K. V. Kucha\v r, \prd 4, 955,
1971.}

\refis{Kuchar1972}{K. V. Kucha\v r, \jmp 13, 758, 1972}.

\refis{Kuchar1976}{K. V. Kucha\v r, \jmp 17, 801, 1976.}

\refis{Kuchar1978}{K. V. Kucha\v r, \jmp 19, 390, 1978.}

\refis{Kuchar1981}{K. V. Kucha\v r, \jmp 22, 2640, 1981.}

\refis{Kuchar1992}{K. V. Kucha\v r, ``Time and Interpretations of
Quantum Gravity'', to
appear in the proceedings of {\it The Fourth Canadian Conference on
General Relativity and Relativistic Astrophysics}, edited by G. Kunstatter,
D. Vincent, and J. Williams (World Scientific, Singapore 1992).}

\refis{Lanczos1970}{C. Lanczos, {\it The Variational Principles of
Mechanics} (University of Toronto Press, Toronto 1970).}

\refis{Lewandowski1992}{J. Goldberg, J. Lewandowski, C. Stornaiolo,
\cmp 148, 377, 1992.}

\refis{Lie1896}{S. Lie, {\it Geometrie der Beruhrungstransformationen},
(B. G. Teubner, Leipzig 1896).}

\refis{Mikhailov1991}{A. Mikhailov, A. Shabat, and V. Sokolov in {\it
What is Integrability?}, ed. V. Zakharov (Springer-Verlag, New York
1991).}

\refis{Noether1918} {E. Noether, \journal Nachr. Konig. Gesell. Wissen.
Gottinger Math. Phys. Kl., , 235, 1918.}

\refis{Olver1986}{P. Olver, {\it Applications of Lie Groups to
Differential Equations}, (Springer-Verlag, New York 1986).}

\refis{Osgood}{See, {\it Conceptual Problems
of Quantum Gravity}, edited by A. Ashtekar and J. Stachel, (Birkh\"auser,
Boston 1991).}

\refis{Ovsiannikov1982}{L. Ovsiannikov, {\it Group Analysis of
Differential Equations}, (Academic Press, New York 1982).}

\refis{Penrose1960}{R. Penrose, \ann 10, 171, 1960.}

\refis{Penrose1976}{R. Penrose, \grg 7, 31, 1976.}

\refis{Penrose1984}{R. Penrose and W. Rindler, {\it Spinors and Space-
Time, Vol. 1}, (Cambridge University Press, Cambridge 1984).}

\refis{Pullin1991}{J. Pullin in {\it Relativity and Gravitation: Classical and
Quantum}, edited by J. C. D'Olivo {\it et al}, (World Scientific, Singapore
1991).}

\refis{Smolin1990}{C. Rovelli and L. Smolin, \np B331, 80, 1990.}

\refis{Sundermeyer1982}{K. Sundermeyer, {\it Constrained Dynamics},
(Springer-Verlag, Berlin 1982).}

\refis{Teitelboim1976}{A. Hanson, T. Regge, C. Teitelboim, {\it
Constrained Hamiltonian Systems}, (Accademia Nazionale dei Lincei,
Rome 1976).}

\refis{Thomas1934}{T. Y. Thomas, {\it Differential Invariants of
Generalized Spaces}, (Cambridge University Press, Cambridge 1934).}

\refis{CGT1989}{K. V. Kucha\v r, C. G. Torre, \jmp 30, 1769, 1989.}

\refis{CGT1990}{K. V. Kucha\v r and C. G. Torre, \prd 43, 419, 1990.}

\refis{CGT1991a}{K. V. Kucha\v r and C. G. Torre, \prd 44,
3116, 1991.}

\refis{CGT1991b}{K. V. Kucha\v r and C. G. Torre in {\it Conceptual
Problems
of Quantum Gravity}, edited by A. Ashtekar and J. Stachel, (Birkh\"auser,
Boston 1991).}

\refis{CGT1991c}{C. G. Torre, \cqg 8, 1895, 1991.}

\refis{CGT1992a}{C. G. Torre, \prd 46, R3231, 1992.}

\refis{CGT1992b}{C. G. Torre, \jmp 33, 3802, 1992.}

\refis{CGT1993}{C. G. Torre  and I. M. Anderson, \prl 70, 3525, 1993.}

\refis{Wald1990}{J. Lee and R. Wald, \jmp 31, 725, 1990.}

\refis{Wald1990a}{R. Wald, \jmp 31, 2378, 1990.}

\refis{Winternitz1989}{C. Boyer and P. Winternitz, \jmp 30, 1081, 1989.}

\refis{Witten1987}{C. Crnkovic and E. Witten in {\it 300 Years of
Gravitation},
edited by S. Hawking and W. Israel (Cambridge University Press,
Cambridge 1987).}

\refis{York1980}{Y. Choquet-Bruhat and J. York in {\it General
Relativity
and Gravitation: 100 Years After the Birth of Albert Einstein, Vol. 1},
edited by A. Held (Plenum, NY 1980).}

\endreferences
\endit